\documentclass[prl,twocolumn,superscriptaddress,fixfloat]{revtex4}

\usepackage{amsmath}
\usepackage{amssymb}
\usepackage{graphicx}
\usepackage{color}  % remove for submission

\begin{document}

\title{Direct observation of paramagnons in palladium}

\author{R. Doubble}
\affiliation{H.H. Wills Physics Laboratory, University of Bristol, Tyndall Ave., Bristol, BS8 1TL, UK}
\author{S. M. Hayden}
\email{s.hayden@bris.ac.uk}
\affiliation{H.H. Wills Physics Laboratory, University of Bristol, Tyndall Ave., Bristol, BS8 1TL, UK}
\author{Pengcheng Dai}
\affiliation{Department of Physics and Astronomy, The University of Tennessee, Knoxville, Tennessee 37996-1200, USA}
\affiliation{Oak Ridge National Laboratory, Oak Ridge, Tennessee 37831, UK}
\author{H. A. Mook}
\affiliation{Oak Ridge National Laboratory, Oak Ridge, Tennessee 37831, UK}
\author{J. R. Thompson}
\affiliation{Department of Physics and Astronomy, The University of Tennessee, Knoxville, Tennessee 37996-1200, USA}
\affiliation{Oak Ridge National Laboratory, Oak Ridge, Tennessee 37831, UK}
\author{C. D. Frost}
\affiliation{ISIS Facility, Rutherford Appleton Laboratory, Chilton, Didcot, Oxfordshire OX11 0QX, UK}


\begin{abstract}
We report an inelastic neutron scattering study of the spin fluctuations in the nearly-ferromagnetic element palladium. Dispersive over-damped collective magnetic excitations or ``paramagnons'' are observed up to 128 meV. We analyze our results in terms of a Moriya-Lonzarich-type spin fluctuation model and estimate the contribution of the spin fluctuations to the low temperature heat capacity. In spite of the paramagnon excitations being relatively strong, their relaxation rates are large. This leads to a small contribution to the low-temperature electronic specific heat.

\end{abstract}

\maketitle

Nearly ferromagnetic metals are of topical interest because their bulk electronic properties can be modified by the presence of spin fluctuations \cite{Doniach1967a,Brinkman1968a,Moriya1985a,Lonzarich1985a,Lonzarich1986a}.
Doniach pointed out \cite{Doniach1967a} that metals close to ferromagetic order at zero temperature should show dispersive overdamped magnetic excitations or ``paramagnons''.
These should be contrasted with the well-defined propagating spin waves which occur in an ordered ferromagnetic phase.  Overdamped modes are still important because they are excited with increasing temperature and therefore contribute to the electronic specific heat \cite{Doniach1967a,Brinkman1968a,Moriya1985a,Lonzarich1985a,Lonzarich1986a}. It has also been suggested that they can mediate superconductive pairing \cite{Foulkes1977a,Fay1977a}.
In this paper, we report an inelastic neutron scattering (INS) study of paramagnons in the element palladium. Pd is unique among the paramagnetic elements in that it shows a large and temperature dependent susceptibility \cite{Kittel}. It has one of the highest densities of states (DOS) \cite{Andersen1970a,Stenzel1986a,Staunton2000a,Larson2004a} at the Fermi energy of the $d$-band metals and the measured susceptibility is approximately 10 times larger than that calculated directly from DOS. Thus it is an good system in which to search for paramagnons. We find that the paramagnon excitations can be observed over a wide range of energy between 25 and 128 meV in the present experiment.

\begin{figure}
\begin{center}
\includegraphics[width=0.8\linewidth]{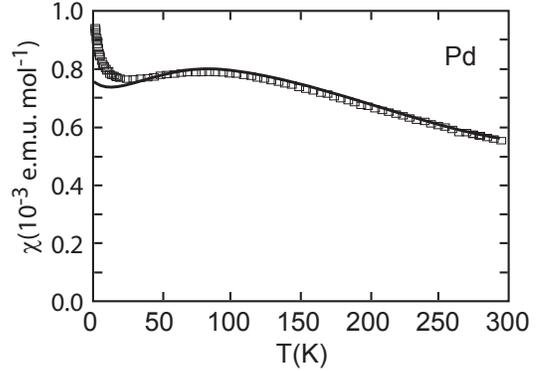}
\end{center}
\caption{The bulk susceptibility of the Pd single crystal used in the present experiment (squares) compared to a standard powder.}
\label{Fig:Pd_chi_bulk}
\end{figure}
Palladium is an face-centered-cubic (FCC) metal with lattice parameter $a$=3.88~\AA.
We studied a 487g single crystal of approximately cylindrical shape with a mosaic of approximately 1.5 deg full-width-at-half-maximum (FWHM). Prior to the experiment the crystal was annealed at a temperature of 300~C under a vacuum of approximately $10^{-6}$ torr for 72h to expel hydrogen \cite{Alefeld1978a}. Fig.~\ref{Fig:Pd_chi_bulk} shows the susceptibility of a piece cut from our sample compared to a powder standard. Both the reference sample and the single crystal used in the experiment show an upturn in the susceptibility at low temperatures.
It is known that even small concentrations of magnetic impurities such as Fe can cause such an upturn at low temperatures [11] due to paramagnetism of the Fe ``giant moments'': based on the magnitude
of the upturn, we estimate the concentration of magnetic impurities to be 60-90 ppm.

INS experiments were performed on the MARI instrument at the ISIS spallation source. MARI is a low-background direct-geometry time-of-flight chopper spectrometer.  For the present experiment, we used detectors located in a single plane henceforth known as the scattering plane. The $(1\bar{1}0)$ crystal plane was mounted coincident with the scattering plane for the present experiment allowing wavevectors of the type $\mathbf{Q}=(h,h,\ell)$ to be investigated.  INS probes the $E$ and $\mathbf{Q}$ dependence of $\chi^{\prime\prime}(\textbf{Q},\omega)$.  The magnetic cross section is given by
\begin{equation}
\label{Eq:cross_sect}
\frac{d^2\sigma}{d\Omega \, dE} = \frac{2(\gamma
r_{\text{e}})^2}{\pi g^{2} \mu^{2}_{\rm B}} \frac{k_f}{k_i} \left| F({\bf Q})\right|^2
\frac{\chi^{\prime\prime}({\bf Q},\hbar\omega)}{1-\exp(-\hbar\omega/kT)},
\end{equation}
where $(\gamma r_{\text{e}})^2$=0.2905 barn sr$^{-1}$, ${\bf k}_{i}$ and ${\bf k}_{f}$ are the incident and final neutron wavevectors and $|F({\bf Q})|^2$ is the magnetic form factor for a Pd 4d orbital \cite{Brown1992}. Data were placed on an absolute scale using a vanadium standard and measurements of the acoustic phonons of the sample. The relatively large size of the Pd crystal meant incident beam was attenuated by absorption and Bragg scattering within the sample. We carried out Monte-Carlo simulations of these effects to account for them in our fitting procedure. We use the reciprocal lattice to label wavevectors $\mathbf{Q}=h\mathbf{a}^{\star}+k\mathbf{b}^{\star}+l\mathbf{c}^{\star}$.

\begin{figure}
\begin{center}
\includegraphics[width=0.8\linewidth]{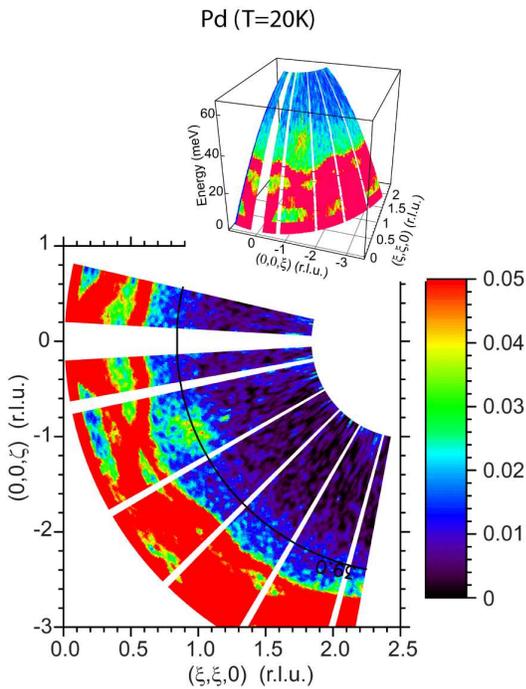}
\end{center}
\caption{(Color online) Paramagnon excitations in Pd at $T$=20~K. Data are collected for a single spectrometer setting with $E_i$=71 meV.  Excitations are probed over the surface of $\mathbf{Q}-\omega$ space shown in the inset. Phonons are observed lower energies $\hbar\omega \lesssim$ 30~meV. The paramagnon scattering can be seen near $\mathbf{Q}$=(1,1,-1), which corresponds to 39~meV.  The black arc is the $\hbar\omega$=39~meV contour.  The units of the plot are mb~St$^{-1}$~meV$^{-1}$~f.u.$^{-1}$.}
\label{Fig:Pd_phoenix_combined}
\end{figure}
\begin{figure}
\begin{center}
\includegraphics[width=0.8\linewidth]{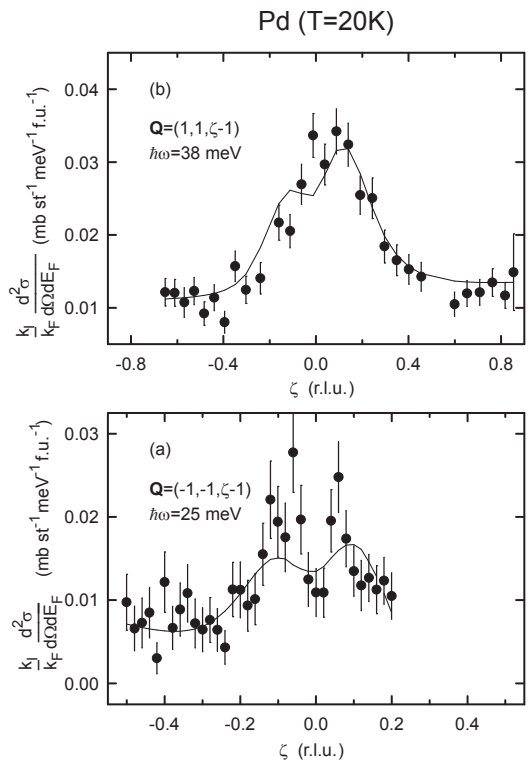}
\end{center}
\caption{Paramagnon excitations at $T$=20~K. Constant energy cuts along $(0,0,\zeta)$ through the (111) reciprocal lattice position for (a) $\hbar\omega$=25~meV and (b) $\hbar\omega$=38~meV. Data were collected with $E_i$=35 and 71 meV respectively. The integrated proton current delivered to the target during the run was 3500 $\mu$Ah.}
\label{Fig:Pd_scans_20K}
\end{figure}
\begin{figure}
\begin{center}
\includegraphics[width=0.8\linewidth]{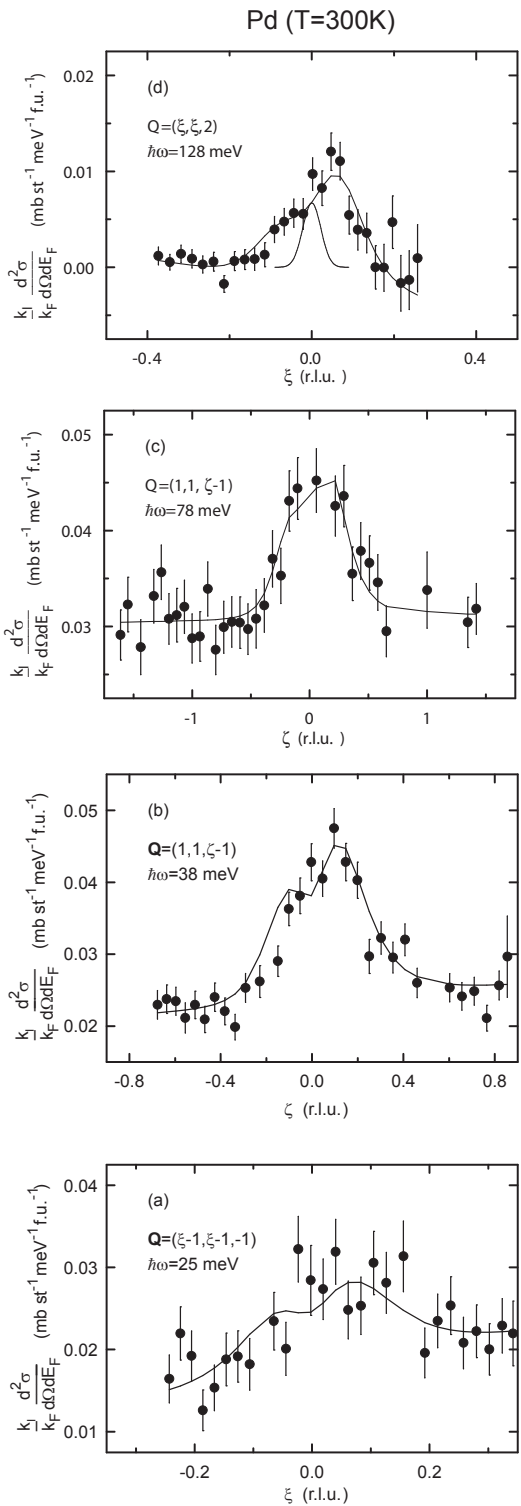}
\end{center}
\caption{Paramagnon excitations at $T$=300~K. Constant energy cuts through the (111) and (200) reciprocal lattice points. Incident energies used were $E_i$=35 (a), 71(b), 204(c) , 300 meV (d). Solid lines are fits to Eq.~\ref{Eq:cross_sect}-\ref{Eq:chi}. The additional line in (d) shows the instrumental response to an energy-independent sharp response $\delta(\mathbf{Q}-\mathbf{\tau})$.}
\label{Fig:Pd_scans_300K}
\end{figure}
In order to avoid any possible complications associated with the low-temperature spin freezing \cite{Peters1984a}, we collected data at $T=20$~K.  Fig.~\ref{Fig:Pd_phoenix_combined} shows data collected with an incident energy $E_i$=71~meV with the [110] direction parallel to $\mathbf{k}_i$.  The main panel shows the scattering function $(k_i/k_f)(d^2\sigma/d\Omega dE)$ plotted as function of wavevector of the excitations $\mathbf{Q}=\mathbf{k}_i-\mathbf{k}_f$. Because the data are collected in a single setting, the energy of the excitations probed $\hbar\omega=E_i-E_f$ varies over the figure. The energy transfer corresponding to each wavevector can be determined from the inset to the figure. At low energies below about 30~meV, we observe the highly structured phonon scattering. Above the highest phonon energy we observe additional scattering near the $\mathbf{Q}=(1,1,-1)$ reciprocal lattice position. For the present setting this corresponds to an energy $\hbar\omega$=39~meV. Fig.~\ref{Fig:Pd_scans_20K}(b) shows a cut directly through this position demonstrating that the additional scattering is peaked at $\mathbf{Q}=(1,1,-1)$.  We note that there is no observable scattering at the $(2,2,-2)$ reciprocal lattice position with larger $|\mathbf{Q}|$. This is consistent with the expected drop in the form factor $\left| F({\bf Q})\right|^2$ by a factor of more than 100 \cite{Brown1992} expected for magnetic scattering. Fig.~\ref{Fig:Pd_scans_20K}(a) shows that the response near the $(11\bar{1})$ zone center is also present at lower energies. Indeed it is peaked at either side of $(11\bar{1})$. Fig.~\ref{Fig:Pd_scans_300K} shows data collected at $T=300$~K for incident energies $E_i=$35.4, 71, 204.3, 300~meV. The scattering near the zone center positions of $(111)$ and $(002)$ types persists up to the highest energies investigated in the present experiment, $\hbar\omega$=128~meV.  It is interesting to note that width of the response is not resolution limited [see Fig.~\ref{Fig:Pd_scans_300K}(d)] and it broadens with increasing energy transfer.

Magnetic excitations in nearly ferromagnetic and weakly magnetic metals are usually interpreted in terms of correlated particle-hole pairs (Stoner excitations).  Doniach \cite{Doniach1967a} calculated the interacting spin-susceptibility for a nearly ferromagnetic metal using the RPA formulation of Izuyama \textit{et al.} \cite{Izuyama1963a}, for a single band Hubbard model. Moriya \cite{Moriya1985a} and Lonzarich \cite{Lonzarich1985a,Lonzarich1986a} (ML) developed a generalized phenomenological form which can be used to describe nearly ferromagnetic metals and provides a useful input to spin-fluctuation theories. In the ML model the imaginary part of generalized susceptibility \cite{note_sus} takes the form,
\begin{equation}
\chi^{\prime\prime}(q,\omega)=\frac{\chi(q) \omega \Gamma(q)}{\Gamma^2(q)+\omega^2},
\label{Eq:Lorz}
\end{equation}
where the relaxation rate $\Gamma(q)$ is given by
\begin{equation}
\Gamma(q)=\gamma q \chi^{-1}(q)
\label{Eq:gamma}
\end{equation}
and the wavevector-dependent susceptibility $\chi(q)=\chi(q,\omega=0)$ is given by:
\begin{equation}
\chi^{-1}(q)=\chi^{-1}+c q^2.
\label{Eq:chi}
\end{equation}
The above expansions are valid in the small $q=|\mathbf{q}|$ limit and $\mathbf{q}$ is measured from a reciprocal lattice point.  Within this model the response is characterized by three microscopic parameters $c$, $\gamma$ and $\chi$, of which only the susceptibility is temperature dependent. 

We first fitted our $T$=300~K data to Eqs.~\ref{Eq:cross_sect}-\ref{Eq:chi}. The solid lines in Fig.~\ref{Fig:Pd_scans_300K} are the results of our fits.  Because the bulk susceptibility of Pd is well known, we fixed this parameter and allowed an overall scale factor to take up any errors in the absolute normalization. We found this factor to be $0.8 \pm 0.2$ i.e. unity within the error of the experiment. The results of fitting the model to the $T$=300~K data are shown in Table \ref{tab:table}.  Within the ML model only the susceptibility is expected to vary significantly with temperature. The lines in Fig.~\ref{Fig:Pd_scans_20K} show the predictions of the model using the $T=20$~K value of the susceptibility from Fig.~\ref{Fig:Pd_chi_bulk}. It is interesting to note that the response at the lowest energy (25 meV) is slightly sharper than the model predicts.  The doubled peaked structure seen in Fig.~\ref{Fig:Pd_scans_20K}(a) is a consequence of the ML model, because $\Gamma(q)\rightarrow 0$ as $q\rightarrow 0$ and does not indicate propagating excitations.
In spite of being a small $q$ and $\omega$ model the phenomenological ML-model appears to provide a reasonable global description of the data.  
The parameters in Table.~\ref{tab:table} can be estimated from electronic structure calculations \cite{Stenzel1986a,Staunton2000a,Larson2004a}. The calculated values are $\hbar \gamma$ = 2.1 \cite{Staunton2000a} and 1.1~$\mu_B^2~\mathrm{meV}^{-1}$ \cite{Larson2004a} and $c$= 836 \cite{Stenzel1986a}, 900 \cite{Staunton2000a} and 925 $\mu_B^{-2}$~\AA$^2$~meV \cite{Larson2004a}. It is not clear why there is a significant discrepancy in the estimation of $c$. However, it has been noted that $c$ is very sensitive to the detailed band structure near the Fermi energy \cite{Staunton2000a,Larson2004a}.

\begin{table}
\begin{center}
\begin{tabular}{|l|c|c|c|c|}\hline
 & $\hbar\gamma$ & $c$ & $\chi^{-1}$ \\
 & $(\mu_B^2$ \AA\ f.u.$^{-1}$) & ($\mu_B^{-2}$ \AA$^2$ meV f.u.) &
  ($\mu_B^{-2}$ meV f.u.) \\ \hline
Pd & $1.74 \pm 0.80$ & $294 \pm 130$ & 41.1 \\ \hline
Ni$_3$Ga & 2.6 & 116 & 2.0 \\ \hline
\end{tabular}
\end{center}
\caption{The results of fitting the phenomenological ML-model (Eqs.~\ref{Eq:Lorz}-\ref{Eq:chi}) to our data \cite{note_sus}. The results  \cite{Bernhoeft1989a} of a similar analysis for Ni$_3$Ga are also given.}
\label{tab:table}
\end{table}

The present results can be compared with those obtained on Ni$_3$Ga. This material is closer to ferromagnetic order at low temperatures and shows a susceptibility enhancement of about 100 with respect to simple band structure calculation \cite{Hayden1986a}. Unfortunately large single crystals of Ni$_3$Ga are not available. Bernhoeft \textit{et al}. \cite{Bernhoeft1989a} carried out an INS study on polycrystalline material at low energies and found that the response could be parameterized in using the model used here. The parameters found are shown in Table~\ref{tab:table}.   The paramagnon excitations in Ni$_3$Ga are observed at much lower energies than in Pd: the maximum energy investigated was a few meV. Surprisingly the $\gamma$ and $c$ parameters in the two materials are the same to within a factor of about 2.  The energy scale of the spin fluctuations is controlled almost entirely by the susceptibility $\chi$ which is different in the two materials. 

As mentioned in the introduction, we would expect \cite{Brinkman1968a,Moriya1985a,Lonzarich1986a,Edwards1992a,Hayden2000a} that the presence of paramagnons will contribute to the low temperature linear specific heat $C=\gamma_{C}T$. We may use our phenomenological response to estimate the contribution of spin fluctuation to the low-temperature specific heat in Pd. Within the ML-model \cite{Lonzarich1986a}, the electronic specific heat has been estimated to be,
\begin{equation}
C=\gamma_{C}T + \delta T^3 \ln (T/T^{\star})
\label{Eq:C_T}
\end{equation}
where
\begin{equation}
 \gamma_{C}=\frac{k_B^2}{4 \pi \hbar \gamma c} \ln(1+c \chi q_u^2),
\end{equation}
 $\delta=2 \pi k_B^4 \chi^3/(5 \hbar^3 \gamma)$, and $T^{\star} \approx \hbar \gamma/(k_B c^{\frac{1}{2}} \chi^{\frac{3}{2}})$.  Using our values for $c$, $\gamma$ and $\chi$ and a cut off wavevector $q_u$ at the Brillouin zone boundary ($q_{\mathrm{BZ}}$=1.71 \AA$^{-1}$), we obtain an estimate of the electronic specific heat of $\gamma_C$=$5.0 \pm 2.7$~mJ K$^{-2}$~mole$^{-1}$. This should be compared with that obtained directly from band structure using the the standard relation
$\gamma_{C}=(\pi^2/3)k_B^2 N(\varepsilon_F)$ and the calculated $N(\varepsilon_F)$=32.7 states atom$^{-1}$ Ry$^{-1}$  \cite{Andersen1970a}, which yields $\gamma_C$=5.6~mJ K$^{-2}$~mole$^{-1}$. We should also note that enhancement of the linear specific heat due electron-phonon coupling is estimated to lie in the range 28-41\% \cite{Papaconstantopoulos1977a,Pinski1978a} for Pd and the experimentally determined value is
$\gamma_C$=9.42~mJ K$^{-2}$ \cite{Veal1964a}.  Combining these facts suggests that the enhancement in the electronic specific heat due to spin fluctuations in Pd is in the range 30--40\% which is within the uncertainty range of our estimate based on the ML-model. Note the estimation based on the ML-model is determined only by experimentally measured quantities. Thus we have a consistent picture in which the observation of strongly spin fluctuations in Pd does not lead to a large contribution to the linear specific heat.

In summary, we have used inelastic neutron scattering to measure so-called ``paramagnon'' excitations in palladium. Paramagnons are dispersing overdamped collective excitations which are present in nearly ferromagnetic metals.  We observe a dispersing response which is strong near the Brillouin zone center and broadens in wavevector with increasing energy up to the highest energies investigated, $\hbar\omega$=128~meV.  We parameterize the observed response and use a Moriya-Lonzarich spin-fluctuation model to estimate the low-temperature linear specific heat directly from our data. We find that relatively small enhancement of the specific heat observed in Pd is consistent with observed paramagnon spectrum which is broad in energy compared to more strongly enhanced systems such as heavy fermions \cite{Aeppli1988a}.

We are grateful to I. I. Mazin for helpful discussions.

%\bibliographystyle{aps2etal}
%\bibliography{Pd_MARI}

\begin{thebibliography}{25}
\bibitem[{not()}]{note_sus}
\bibinfo{note}{Susceptibilities in this paper are generally quoted in units of
  $\mu_B^2$~meV$^{-1}$, defined through the equation for the Pauli
  susceptibility of a metal $\chi=\mu_B^2 N(\varepsilon_F)$ or the energy of a
  paramagnet in a magnetic field $E=M^2/(2 \chi)$.}
\end{thebibliography}

\end{document}